\documentclass[aps,prb,twocolumn,superscriptaddress,showkeys]{revtex4-1}

\usepackage{amsmath}
\usepackage{amssymb}
\usepackage{graphicx}

\begin{document}

\title{Nonlocal study of ultimate plasmon hybridization}
\date{\today}

\author{S\o ren Raza}
\affiliation{Department of Photonics Engineering, Technical University of Denmark, DK-2800 Kgs. Lyngby, Denmark}
\affiliation{Center for Electron Nanoscopy, Technical University of Denmark, DK-2800 Kgs. Lyngby, Denmark}
\author{Martijn Wubs}
\affiliation{Department of Photonics Engineering, Technical University of Denmark, DK-2800 Kgs. Lyngby, Denmark}
\affiliation{Center for Nanostructured Graphene (CNG), Technical University of Denmark, DK-2800 Kgs. Lyngby, Denmark}
\author{Sergey I. Bozhevolnyi}
\affiliation{Department of Technology and Innovation, University of Southern Denmark, DK-5230 Odense, Denmark}
\author{N. Asger Mortensen}
\affiliation{Department of Photonics Engineering, Technical University of Denmark, DK-2800 Kgs. Lyngby, Denmark}
\affiliation{Center for Nanostructured Graphene (CNG), Technical University of Denmark, DK-2800 Kgs. Lyngby, Denmark}
\email{Corresponding author: asger@mailaps.org}

\begin{abstract}
Within our recently proposed generalized nonlocal optical response (GNOR) model, we revisit the fundamental problem of an optically excited plasmonic dimer. The dimer consists of two identical cylinders separated by a nanometre-sized gap. We consider the transition from separated dimers via touching dimers to finally overlapping dimers. In particular, we focus on the touching case, showing a fundamental limit on the hybridization of the bonding plasmon modes due to nonlocality. Using transformation optics we determine a simple analytical equation for the resonance energies of the bonding plasmon modes of the touching dimer.
\end{abstract}

\maketitle

\section{Introduction}
One of the most fundamental and intriguing problems in plasmonics is the electromagnetic interaction of two metallic nanostructures, i.e., the dimer structure. The gap-dependent electric-field enhancement and bonding plasmon resonance energies have been utilized in e.g. surface-enhanced Raman spectroscopy~\cite{Kneipp:2007} and the plasmon ruler effect~\cite{Jain:2007}. The dimer has been studied with a variety of theoretical and experimental techniques. The simplest theoretical approach is based on the classical local-response approximation (LRA), which in the extreme case of a nanometre-sized dimer gap gives rise to unphysical results, such as diverging field enhancements in the gap of the dimer~\cite{Romero:2006}. A complete breakdown of the LRA is seen in the touching configuration, where the number of hybridized bonding plasmon modes increases without bound to form a continuum of modes~\cite{Fernandez-Dominguez:2012,Fernandez-Dominguez:2012b}. Advanced descriptions based on density-functional theory (DFT) regularize the unphysical consequences of the LRA~\cite{Zuloaga:2009a,Stella:2013,Teperik:2013b,Andersen:2013}, where the physical mechanism for the regularization is attributed to quantum tunnelling, i.e., the charge transfer that may occur before reaching the touching configuration due to the spill-out of electrons. Numerical DFT calculations of optically excited dimers show a limit on the hybridization of bonding plasmon modes, yet no general relation for this limit has been extracted. Results based on the hydrodynamic model, where only nonlocal response (and not electron spill-out) is taken into account, also display regularizations of the LRA~\cite{Fernandez-Dominguez:2012,Fernandez-Dominguez:2012b,Toscano:2012a}, albeit with field enhancements in the dimer gap that are still larger than shown by DFT simulations~\cite{Teperik:2013b}. Measurements on dimers with vanishing gaps using both optical techniques~\cite{Savage:2012} and electron energy-loss spectroscopy~\cite{Scholl:2013,Kadkhodazadeh:2013} are not in agreement with the LRA, and, in the touching case, also display limits on the resonance energies of the bonding plasmon modes. However, the physical mechanism for the discrepancy between LRA and the observed measurements is not conclusive.

\begin{figure}[!b]
\center
\includegraphics[scale=1]{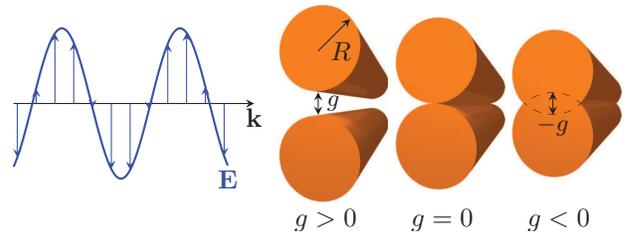}
\caption{Sketch of the considered system, displaying the incident electric field, which is polarized along the dimer axis, impinging on a dimer consisting of two identical cylinders with radii $R$ and separated by a gap $g$. The three cases of separated dimer ($g>0$), touching dimer ($g=0$) and overlapping dimer ($g<0$) are shown.}
\label{fig:fig1}
\end{figure}

In this paper, we revisit the problem of a plasmonic dimer, which consists of two identical cylinders excited by a plane wave (see Fig.~\ref{fig:fig1}), within the framework of our recently proposed GNOR model in which nonlocal response is caused both by convection and diffusion~\cite{Mortensen:2014}. In contrast to DFT, the physically more transparent GNOR model allows for analytical results. We study the evolution of both the extinction cross section and the electric-field enhancement in the dimer gap as the gap size varies from separated to overlapping dimers. In particular, we focus on the touching dimer and derive, using the formalism of transformation optics (TO), a simple analytical relation for the resonance energies of the bonding plasmon modes. Furthermore, we show that previous nonlocal TO methods used for the hydrodynamic model~\cite{Fernandez-Dominguez:2012,Fernandez-Dominguez:2012b} are accurate also for the GNOR model.

\section{Generalized nonlocal optical response}
Given the nonlocal constitutive relation connecting the displacement field with the electric field $\mathbf D(\mathbf r,\omega) = \int \text d \mathbf{r'} \varepsilon(\mathbf r,\mathbf{r'}) \mathbf E(\mathbf{r'},\omega)$, the wave equation is in general given as
\begin{equation}
    \mathbf{\nabla} \times \mathbf{\nabla} \times \mathbf E(\mathbf r,\omega) = \left(\frac{\omega}{c}\right)^2 \int \text d \mathbf{r'} \varepsilon(\mathbf r,\mathbf{r'}) \mathbf E(\mathbf{r'},\omega). \label{eq:generalnl}
\end{equation}
Under the assumptions of a linear, isotropic and short-ranged response function, Eq.~(\ref{eq:generalnl}) can be recast as~\cite{Mortensen:2013}
\begin{equation}
    \mathbf{\nabla} \times \mathbf{\nabla} \times \mathbf E(\mathbf r,\omega) = \left(\frac{\omega}{c}\right)^2 \left[\varepsilon_\textsc{d}(\omega) + \xi^2 \nabla^2 \right] \mathbf E(\mathbf r,\omega), \label{eq:nl}
\end{equation}
where the parameter $\xi$ describes the nonlocal correction to the local-response Drude permittivity $\varepsilon_\textsc{d}(\omega)=\varepsilon_\infty(\omega)-\omega_\textsc{p}^2/(\omega^2+i\gamma\omega)$. Here, $\varepsilon_\infty(\omega)$ accounts for effects not due to the free electrons, such as interband transitions. Within the nonlocal hydrodynamic model~\cite{Boardman:1982a}, the nonlocality parameter is $\xi_\textsc{h}^2 = \beta^2/(\omega^2+i\omega\gamma)$~\cite{Toscano:2013}, where $\gamma$ is the Drude free-electron damping and $\beta^2 = 3/5 v_\textsc{f}^2$ with $v_\textsc{f}$ denoting the Fermi velocity of the conduction electrons. Nonlocal response in the hydrodynamic model arises from the inclusion of the Thomas--Fermi kinetic energy of the free electrons. The GNOR model incorporates the hydrodynamic approach and expands it by taking into account the effects of electron diffusion~\cite{Mortensen:2014}. The nonlocality parameter within the GNOR model is $\xi_\textsc{gnor}^2 = \eta^2/(\omega^2+i\omega\gamma)$, where $\eta^2=\beta^2+D(\gamma-i\omega)$ and $D$ is the diffusion constant. Although the mathematical formalism of the GNOR model is similar to the hydrodynamic model, with the simple substitution $\beta^2 \rightarrow \eta^2$, the physical consequences are pronounced. In contrast to the hydrodynamic model, the GNOR model accurately captures the size-dependent damping of localized surface plasmons in individual particles~\cite{Kreibig:1995} and reproduces time-dependent DFT absorption spectra of dimers~\cite{Teperik:2013b,Mortensen:2014}.

\begin{table}[!b]
\begin{tabular}{lccccc}
	\hline
    & $\hbar \omega_\textsc{p}$ [eV] & $\hbar \gamma$ [eV] & $v_\textsc{f}$ [$10^6$ms$^{-1}$] & \multicolumn{2}{c}{$D$ [$10^{-4}$ m$^2$s$^{-1}$]} \\
    \cline{5-6}
    & & & & $A=0.5$ & $A=1$ \\
    \hline
    Na & 6.04~\cite{Ashcroft:1976} & 0.16~\cite{Teperik:2013b} & $1.07$ & $1.08$ & $2.67$ \\
    Ag & 8.99~\cite{Ashcroft:1976} & 0.025~\cite{Abajo:2010} & $1.39$ & $3.61$ & $9.62$ \\
    Au & 9.02~\cite{Ashcroft:1976} & 0.071~\cite{Abajo:2010} & $1.39$ & $1.90$ & $8.62$ \\
    Al & 15.8~\cite{Ashcroft:1976} & 0.6~\cite{Abajo:2010} & $2.03$ & $1.86$ & $4.59$ \\
    \hline
\end{tabular}
\caption{Plasma frequencies $\omega_\textsc{p}$, Drude damping rates $\gamma$, Fermi velocities $v_\textsc{f}$ and diffusion constants $D$ for the metals Na, Ag, Au and Al. The method used for determining $D$ is described in the main text. }
\label{tab:tab1}
\end{table}

Before discussing the results of the nanowire dimer, we first outline the procedure to determine the diffusion constant $D$ for different metals. The size-dependent damping of localized surface plasmons in nanoparticles is well-known and has extensively been observed experimentally~\cite{Kreibig:1995,Kreibig:1969}. The phenomenological approach to account for this effect in the LRA, i.e., Eq.~(\ref{eq:nl}) with $\xi=0$, has been to modify the Drude damping parameter as $\gamma \rightarrow \gamma + A v_\textsc{f}/R$~\cite{Kreibig:1969}, which only applies for spherical particles of radius $R$. Here, $A$ is a constant, which is related to the probability of the electron scattering off the surface of the particle. Both experimental data and more advanced theoretical calculations have been compared to this approach, resulting in a robust value for $A$ close to unity for different metals~\cite{Kreibig:1995}. It is therefore appropriate to ensure that the GNOR model agrees with this successful, but phenomelogical approach. To this end, we calculate the extinction cross section of a metal sphere using the nonlocal polarizability~\cite{Raza:2013c} for the GNOR calculations, while the LRA polarizability is modified to include the aforementioned additional damping with both $A=0.5$ and $A=1$. Two different values for $A$ are considered, since different nanoparticle preparation methods may result in different surface properties. The diffusion constant $D$ is varied until the full-width at half maximum of the localized surface plasmon resonances for both calculations coincide. This procedure is repeated for the range of sphere radii of $1-10$~nm. As the fitted diffusion constant $D$ varies slightly with sphere radii, we use the average value for $D$. The diffusion constant $D$ along with other relevant GNOR parameters for Na, Ag, Au, and Al are summarized in Table~\ref{tab:tab1}. For completeness, we add that the non free-electron response of metals $\varepsilon_\infty(\omega)$ can be determined from experimentally measured bulk dielectric functions $\varepsilon_\text{exp}(\omega)$ using the procedure $\varepsilon_\infty(\omega) = \varepsilon_\text{exp}(\omega)+\omega_\textsc{p}^2/(\omega^2+i\gamma\omega)$~\cite{Abajo:2008}.

\section{Results}

\begin{figure}[!b]
\center
\includegraphics[scale=1]{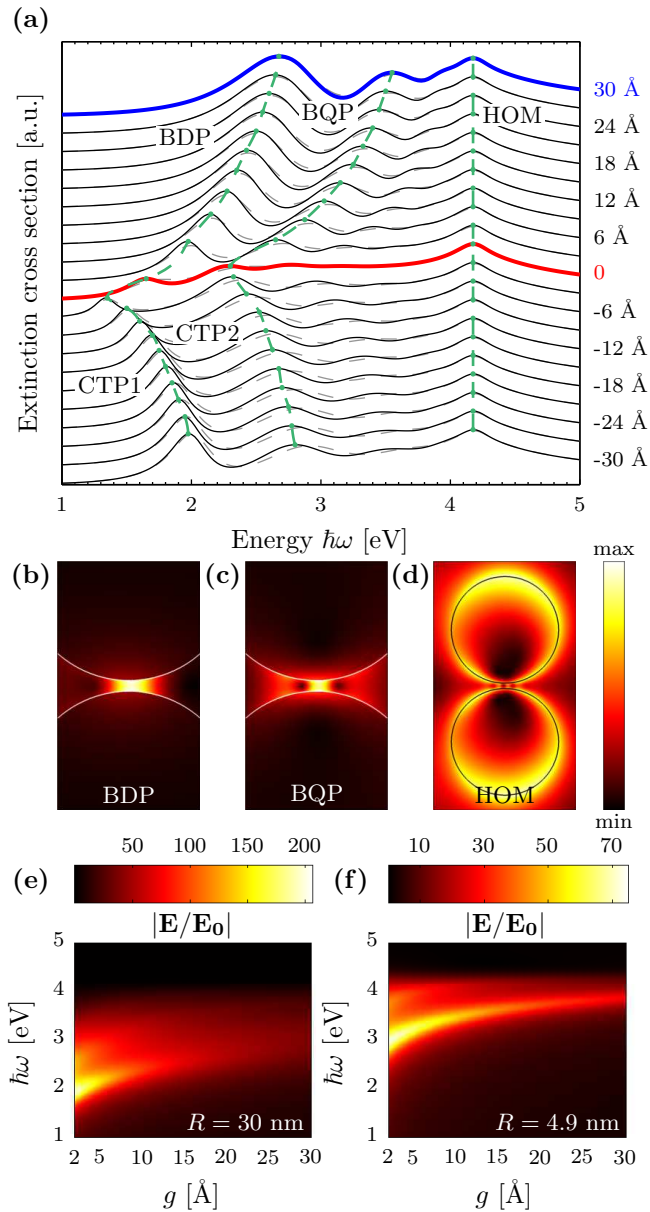}
\caption{\textbf{(a)} Waterfall plot of the extinction cross section of a Na nanowire dimer of radius $R=30$~nm with gap size $g$ varying from $30$~{\AA} (separated) to $-30$~{\AA} (overlapping) for two different diffusion constant values: $D=1.08\times10^{-4}$~m$^2$s$^{-1}$ (solid lines) and $D=2.67\times10^{-4}$~m$^2$s$^{-1}$ (dashed lines). The labels denote the excited modes in the spectra. The maximal hybridization is achieved for the touching dimer (red curve). \textbf{(b-d)} Plots of electric field amplitude $|\mathbf{E}|$ of the BDP, BQP, and HOM, respectively, for a dimer with $g=30$~{\AA} and $D=1.08\times10^{-4}$~m$^2$s$^{-1}$ [blue curve in (a)]. \textbf{(e-f)} Electric-field enhancement in the center of the dimer gap for dimers with $R=30$~nm and $R=4.9$~nm, respectively ($D=1.08\times10^{-4}$~m$^2$s$^{-1}$). Parameter values for Na as in Table~\ref{tab:tab1} with $\varepsilon_\infty=1$.}
\label{fig:fig2}
\end{figure}

Using the freely available COMSOL implementation of the GNOR theory~\cite{Toscano:2012a} we calculate the extinction cross section and electric-field enhancement of a large dimer with radius $R=30$~nm when varying the gap distance. Figure~\ref{fig:fig2}(a) displays a waterfall plot of the extinction cross section for two values of $D$ (solid and dashed lines, respectively), when the gap is varied from $+30$~{\AA} (separated) to $-30$~{\AA} (overlapping) in steps of $3$~{\AA}. For separated nanowires, we see the presence of three distinct modes, the bonding dipole mode (BDP), bonding quadrupolar mode (BQP) and the higher-order mode (HOM), also labeled in Fig.~\ref{fig:fig2}. The electric field norm of these three modes for a gap of $30$~{\AA} are displayed in Fig.~\ref{fig:fig2}(b-d). As the gap decreases, the BDP and BQP redshift and dampen due to increased plasmon hybridization~\cite{Fernandez-Dominguez:2012b,Toscano:2012a} and increased absorption at the metal surfaces in the vicinity of the gap, respectively. However, the resonance energy of the HOM is unaffected by the gap size as the electric field of this mode is mainly distributed at the dimer edges, and not in the gap [see Fig.~\ref{fig:fig2}(d)]. When the gap vanishes [red line in Fig.~\ref{fig:fig2}(a)] the hybridization of the bonding modes is maximal, and no further resonance shifts or damping can occur. As the dimers begin to overlap ($g<0$) the bonding modes disappear and the charge transfer plasmons [labeled CTP1 and CTP2 in Fig.~\ref{fig:fig2}(a)] appear. The CTP1 and CTP2 blueshift with increasing dimer overlap as the dimer effectively becomes a single structure. Comparison of the two values for $D$ [solid and dashed lines in Fig.~\ref{fig:fig2}(a)] reveal that the main features are robust with respect to $D$, and that an increased value for $D$ primarily increases the widths of the bonding-mode resonances. In Fig.~\ref{fig:fig2}(e) we show the evolution of the electric-field enhancement, probed at the center point of the gap. As the gap size decreases, the electric-field enhancement increases due to increased interaction between the metal surfaces. Furthermore, as the bonding plasmon modes redshift so does the maximal field enhancement. To facilitate comparison with DFT calculations~\cite{Teperik:2013b} and other theoretical models based on quantum tunnelling, such as the quantum-corrected model (QCM)~\cite{Esteban:2012} and the quantum conductivity theory (QCT)~\cite{Haus:2014}, we also display in Fig.~\ref{fig:fig2}(f) the electric-field enhancement evolution of a Na dimer with a smaller radius of $R=4.9$~nm. Here we see that the electric-field enhancement amplitude and trend with decreasing gap size are in very good agreement with DFT simulations of Ref.~\cite{Teperik:2013b}. The only discrepancy between the GNOR model and DFT simulations occurs at gap sizes below approximately 5~{\AA} but before contact, where the overlap of electron spill-out in DFT calculations quenches the electric-field enhancement.

\begin{figure}[!t]
\center
\includegraphics[scale=1]{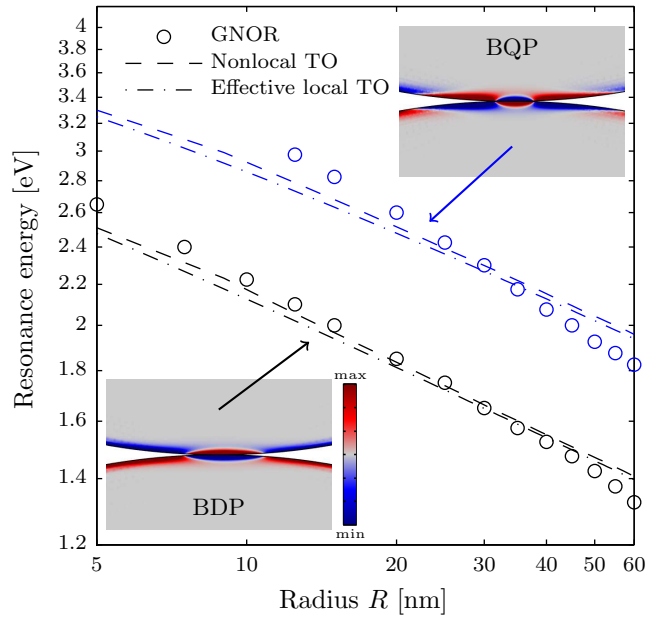}
\caption{Resonance energies of the BDP (black) and BQP (blue) modes of touching Na nanowires ($g=0$) as a function of dimer radius $R$. The dots display the simulations using the GNOR model, and the dashed lines show the results using the nonlocal TO approach. Finally, the dash-dotted lines show the results using the local-response TO approach, given by Eq.~(\ref{eq:LRArescond}) with an effective gap $g=2\text{Re}[d(\omega)]$. The insets display the real part of the GNOR charge distributions of the BDP and BQP for a touching dimer with $R=30$~nm.}
\label{fig:fig3}
\end{figure}

When the dimers are touching, the hybridization of the bonding plasmon modes is maximal and the resonance positions of these  modes depend only on the dimer radius $R$. We have investigated the behavior of the resonance condition of the BDP and BQP modes as a function of $R$ in Fig.~\ref{fig:fig3}. For the smallest dimer radii ($R\leq10$~nm), the resonance positions of the BQP mode are not clearly distinguishable from the extinction spectra due to the weaker hybridization in smaller dimers. As the dimer radius increases the resonance energies of both the BDP and BQP modes decrease. This is due to the increased hybridization occurring for larger radii as the interacting metal surfaces between the two nanowires increase. Along with the numerical GNOR simulations (dots in Fig.~\ref{fig:fig3}), we also depict the results using a nonlocal transformation optics (TO) approach (dashed line)~\cite{Fernandez-Dominguez:2012}. Although the nonlocal TO was originally used with the hydrodynamic model, we show in Fig.~\ref{fig:fig3} that the nonlocal TO approach is still valid within the GNOR theory, as long as the substitution $\beta^2\rightarrow \eta^2$ is applied. As anticipated, we see in Fig.~\ref{fig:fig3} that the nonlocal TO calculations agree quite well with the GNOR simulations for both the BDP and BQP modes.

We may deduce a simple relation for the resonance energies of the bonding plasmon modes by examining the position of the centroid of induced charges~\cite{Teperik:2013b}, given as the real part of the Feibelman parameter $d(\omega)$~\cite{Feibelman:1982}. Using the definition of the Feibelman parameter~\cite{Feibelman:1982}, we find that $d(\omega)= i/k_\textsc{nl}$ where $k_\textsc{nl}=\sqrt{\varepsilon_\textsc{d}(\omega)/\varepsilon_\infty(\omega)} /\xi_\textsc{gnor}$ is the nonlocal longitudinal wave vector. In the GNOR theory, the centroid of the induced charges is positioned a short distance \{$\text{Re}[d(\omega)] \simeq v_\textsc{f}/\omega_\textsc{p} \approx 1$~{\AA}\} within the metal boundary (as a consequence of the additional boundary condition of vanishing free-electron current in the radial direction~\cite{Mortensen:2014,Boardman:1982a}). However, within the LRA the induced surface charges reside on the geometrical surface. We can therefore mimic the position of the centroid of induced charges in the GNOR theory by considering separated dimers in the LRA with a gap of $g=2\text{Re}[d(\omega)]$. Within the LRA, the resonance condition of a separated dimer with gap size $g$ has been determined using TO~\cite{Aubry:2010} and is given by the relation
\begin{equation}
    \left(\sqrt{\frac{g}{4R}} + \sqrt{1+\frac{g}{4R}} \right)^{4n} = \text{Re}\left[\frac{\varepsilon_\textsc{d}(\omega)-1}{\varepsilon_\textsc{d}(\omega)+1}\right], \label{eq:LRArescond}
\end{equation}
where $n=1$ corresponds to the BDP mode, $n=2$ corresponds to the BQP mode and so on. Assuming an undamped Drude model for the permittivity $\varepsilon_\textsc{d}(\omega)=1-\omega_\text{p}^2/\omega^2$ and expanding Eq.~(\ref{eq:LRArescond}) to first-order in $g/R$, we find the simple relation for the LRA resonance condition for the modes of separated nanowires
\begin{equation}
    \frac{\omega}{\omega_\text{p}} \simeq \frac{1}{\sqrt{n}} \left[ \frac{\sqrt{-\varepsilon_\textsc{d}(\omega)} R}{2 \text{Re}(\xi_\textsc{gnor})}  \right]^{\frac 14}, \label{eq:LRAresapprox}
\end{equation}
where we have used that $g=2\text{Re}(\xi_\textsc{gnor})/\sqrt{-\varepsilon_\textsc{d}(\omega)}$. Figure~\ref{fig:fig3} displays the result of this effective LRA approach (dash-dotted lines), given by Eq.~(\ref{eq:LRArescond}) with gap size $g=2\text{Re}[d(\omega)]$. We see that the GNOR resonance energies of touching nanowires can quite accurately be mimicked by the LRA result of separated nanowires, when the gap size is set to the distance between the centroid of induced charges. As anticipated from Eq.~(\ref{eq:LRAresapprox}), we also see that the slope of the BDP and BQP resonance energies are very similar. The BQP energies simply occur at higher energies. Although diffusion plays a crucial role in the damping of the bonding plasmon modes for decreasing gap size [as seen in the extinction cross sections of Fig.~\ref{fig:fig2}(a)] and in the electric-field enhancement amplitude [see Figs.~\ref{fig:fig2}(e-f)], Eq.~(\ref{eq:LRAresapprox}) shows that the maximal hybridization resonance energies are mainly dependent on convection as described by the Fermi velocity, since the value for $\beta$ contributes most to $\text{Re}(\xi_\textsc{gnor})$. Only in the extreme limit where $D\omega$ becomes comparable to $\beta^2$ in magnitude will diffusion play a role in the position of the resonance energies.

\section{Conclusions}
We have theoretically studied the extinction cross section and electric-field enhancement of a plasmonic dimer consisting of two large ($R=30$~nm) Na cylinders using the GNOR model. As the gap size decreases, the extinction cross section shows a damping of the bonding plasmon resonances, while the electric-field enhancement in the gap increases, but stays finite. Both trends are in good agreement with DFT calculations and experimental measurements on dimers.

We have also examined the touching dimer and, using transformation optics, derived a simple analytical relation for the resonance energies of the bonding plasmon modes that we propose to test experimentally. Finally, we have shown the first successful application of nonlocal TO to the GNOR model.

\section*{Funding Information}
The Center for Nanostructured Graphene is funded by the Danish National Research Foundation, Project DNRF58. N.A.M. and M.W. acknowledge financial support by Danish Council for Independent Research—Natural Sciences, Project 1323-00087.

\section*{Acknowledgments}
We thank Yu Luo for stimulating discussions.


\end{document}